# ATLASrift - a Virtual Reality application

Ilija Vukotic[1*], Edward Moyse[2], Riccardo Maria Bianchi[3]

[1]The Enrico Fermi Institute, The University of Chicago, US
[2]University of Massachusetts, US
[3]University of Pittsburgh, US

We present ATLASrift - a Virtual Reality application that provides an interactive, immersive visit to ATLAS experiment. We envision it being used in two different ways: first as an educational and outreach tool - for schools, universities, museums and interested individuals, and secondly as an event viewer for ATLAS physicists – for them it will provide a much better spatial awareness of an event, track and jet directions, occupancies and interactions with detector structures. Using it, one can learn about the experiment as a whole, visit individual sub-detectors, view real interactions, or take a scripted walkthrough explaining questions physicists are trying to answer. We briefly describe our platform of choice – OculusRift VR system, the development environment – UnrealEngine, and, in detail, the numerous technically demanding requirements that had to be fulfilled in order to provide a comfortable user experience. Plans for future versions include making the experience social by adding multi-user/virtual presence options, event animation, interactive virtual demonstrations of key high energy physics concepts, or detector operating principles.

---

* On behalf of the ATLAS outreach group

## 1. MOTIVATION

During the last two decades graphics programs used to visualize high energy physics events changed in so far as to support higher resolutions. They are still very much tied to an experiments software stack, which is used to read and decode events from the input files, load experiment geometry, etc. This also limits them to the most often used computing platform in high energy physics – Linux servers with OpenGL graphics being shown through X window system. Consequently performance is usually bellow one frame per second (FPS).

At the same time commercial graphics programs of which the largest subset are video games, offer more than 100 Hz frame rate, while at the same time handling physics (mechanics of the game), sound, artificial intelligence, and provide multi-user experience over the internet, all on hardware platforms (including even mobile phones) costing only a fraction of the typical HEP server. The latest trend in the field of video games and digital media in general is Virtual Reality (VR).

Virtual Reality replicates an environment that simulates physical presence in places in the real world or imagined worlds. VR has been considered a Holly Grail of computer visualization. Despite a lot of efforts expended in last 40 years, VR systems were plagued by: high latency, low resolution, bulk, price etc. thus limiting their application to a very high profile, custom developed applications e.g. military simulators. Numerous technical developments (lite, high resolution, and low persistency mobile screens, powerful CPUs and GPUs) in the last several years finally solved majority of problems and brought VR to a verge of a mass market with a possibility of making it a new default media consumption platform.

## 2. ATLASRIFT

Virtual reality as a platform is almost ideal for the visualization of high-energy particle collisions, as it enables a physicist to easily grasp events 3D topology, investigates paths and detector volumes crossed by the resulting particles. As there is no support for the VR in the software stack used by the current even viewer software, we decided to break away from all the experiment specific frameworks and obsolete technologies and start from the scratch. Not being tied to experiment software makes the application much easier to create, deliver and install while covering more platforms.

The ATLASrift application [1] has two main use cases: as an event viewer for ATLAS [2] physicists (to them it will provide a much better spatial awareness of an event, track and jet directions, occupancies and interactions with detector structures) , and as an educational and outreach tool - for schools, universities, museums and any interested individuals. To people not in the field the ATLASrift Virtual Reality application provides an interactive, immersive visit to the ATLAS experiment. Using it, one can learn about the experiment as a whole, inspect individual sub-detectors, view real interactions, or take a scripted walkthrough explaining questions physicists are trying to answer.

## 3. TECHNOLOGY CHOICES

Having an exciting scientific content is only one part of the successful VR application. Equally important is to ensure a user does not get nauseated and have as immersive as possible experience usually described as feeling "presence". Some of best practices to achieve this are: more than 75Hz frame rate (ideally 100 Hz), less than 15 ms latency, no elements fixed to a point in the users field of view, realistic movement speeds, etc.

## 3.1. Computing Platform

There are already commercially available VR platforms (Samsung Gear, Google cardboard) and several more still in development phase (Oculus Rift, SONY Morpheus, HTC ReVive, Razer OSVR, SteamVR). We chose to develop for the Oculus Rift [3] as the first and most developed platform even the production version will be available only in 2016. In a meanwhile we use developer kit DK2 with a Microsoft XBOX One controller as a preferred input device.

## 3.2. Software Platform

In house creating an application that supports VR hardware, is sufficiently fast, handles synchronization of graphic rendering, audio, user input, networking etc. is not an option due to complexity, manpower, and time needed. Luckily we can use the same tools used by the multi-billion game industry. There are several commercial game engines that enormously simplify application development, are open source, and free for an educational purposes. Two most popular are Unreal Engine [4] and Unity [5]. We have chosen the former. While it is possible to program everything in C++, programming using a graphical interface called "Blueprints" proved very simple and powerful enough for the task at hand. As long as no platform specific third party libraries are used and the C++ code is written in a platform agnostic way, Unreal Engine provides a single-click project build for Windows, XBOX, Linux, MacOS, PlayStation, iOS, and Android. For the first two it uses DirectX versions 11 or 12, and for the rest OpenGL.

The most development effort went into transforming the detector geometry into a form adequate for the application, creation of user interfaces, walkthrough scripts and accompanying audio.

## 3.3. Content

Application resources may be roughly split into three main groups: geometry, events, and media. Here we briefly describe all three.

### 3.3.1. Geometry
There are two main geometry resources: the detector and its surroundings. The ATLAS detector geometry used by the ATLAS software framework – ATHENA, comes from a geometry database files. The official ATLAS event viewer VP1 [6] was used to export the geometry of all of the sub-detectors separately in a SGI Inventor format. These are then exported into VRML 2.0 format. Obtained 3D models have a very high polygon count so we used Blender [7] and 3DS MAX[8], to simplify it, clean it from unreferenced vertices, create UV meshes, and finally triangulate the models before export in FBX format supported by the Unreal Engine. The 3D models of the detector cavern, counting rooms, walkways, shafts, and supporting structures where obtained from CERN engineering department as CATIA models and passed a very similar import procedure.

### 3.3.2. Events
Several different data format files are used in ATLAS. All of them are ROOT [9] files. Most details of the interaction are preserved in ESD files that contain not only reconstructed objects (vertices, tracks, clusters, etc.) but also hits. xAOD format on the other hand has only reconstructed information. In order to avoid dependency on both ROOT and ATLAS IO libraries, and also avoid a need to deliver the event files with the application, we decided to export the information needed for the visualization in JSON format and index it in our ElasticSearch [10] cluster. This way hundreds, possibly thousands of people can access the events of their choosing concurrently and with minimal lag. Before publishing all the events have to be approved for publication by the ATLAS collaboration. Relaxing this requirement in the future would open the way for displaying a continuous stream of events of the desired type as soon as they get reconstructed.

### 3.3.3. Media

Outreach oriented ATLASrift content consists of both audio and visual media. Both full detector and each sub-detector separately are covered with a number of photos from the CERN photo library, followed by a narration recorded by the ATLAS physicists and ATLAS outreach group. These explain what ATLAS is, what we are trying to measure, and some basics related physics concepts.

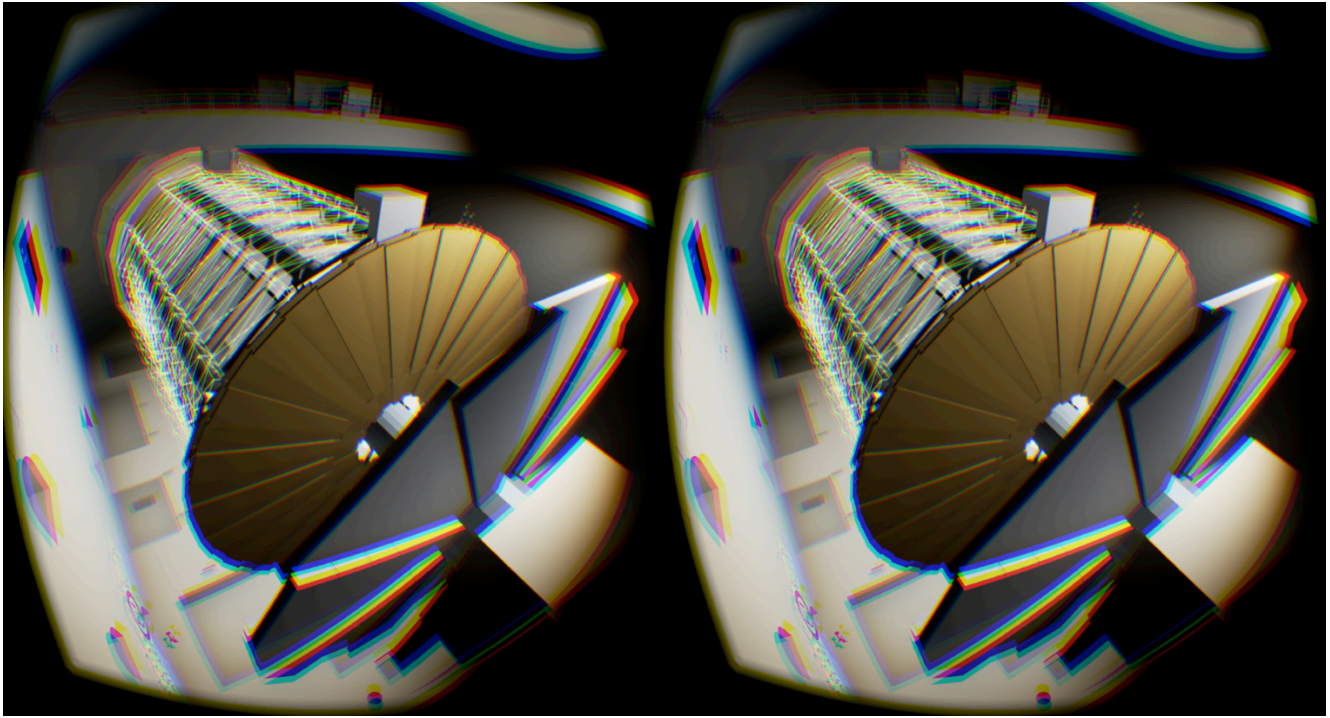

**Figure 1** A stereoscopic view of the ATLAS detector and cavern. Outer muon tracker chambers are shown as a wireframe. Walkways and support structures not shown.

## 4. CHALLENGES OF THE NEW MEDIUM

Except technical challenges (high frame rate, low latency, etc.) a number of issues specific to VR have to be addressed. The first issue was that if we make 3D model of character representing user to be blocked by the objects in the virtual environment, movement around the detector and investigation of details of events becomes almost impossible. Consequently, we chose to have body-less character, not limited by the physics laws (no gravity, can pass through walls). It is difficult to show a lot of small text in VR so hundreds of option available in standard event viewers had to be drastically reduced. Interactions with the menu in 3D are also impossible with the computer mouse and keyboard. Following extensive testing of what works best, we settled on a menu where movement between different options is tied to the head movements (pitch and yawn) while effectuation of thus selected option is done using controller/mouse.

## 5. FUTURE

The initial application has been developed as an R&D project. A large community interest in a project [Figure 2] warrants further development. With a dedicated effort the future versions will include: making the experience social by adding multi-user/virtual presence options, event animation, interactive virtual demonstrations of the key high energy physics concepts, and detector operating principles.

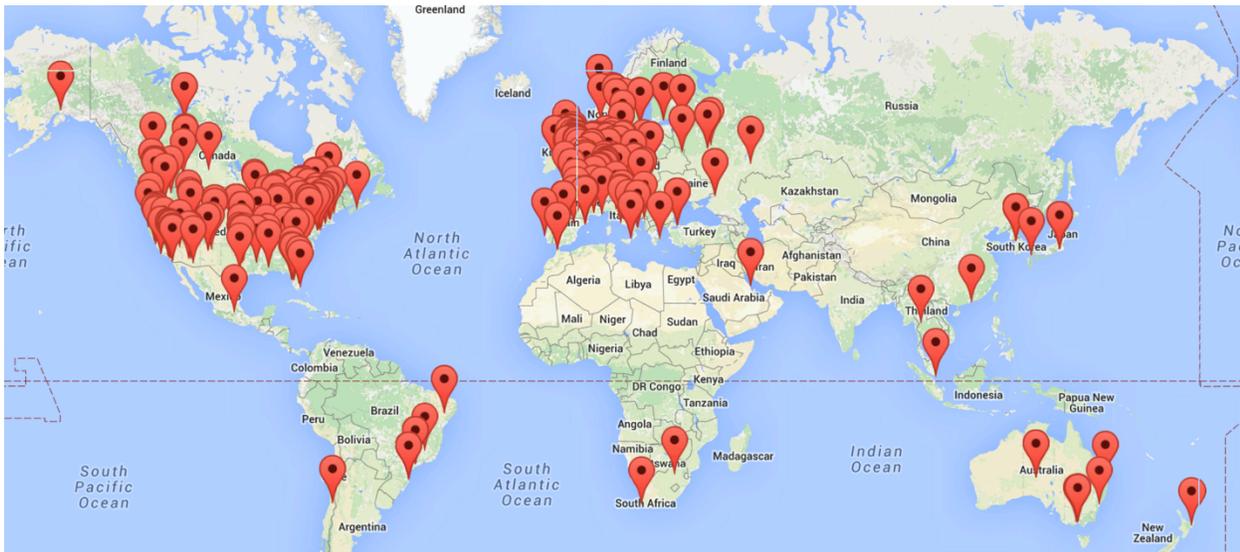

**Figure 2** In just first 20 days the application was available for download, more than 500 people all over the world used it to virtually visit ATLAS experiment.